\documentclass{appolb}
\usepackage{graphicx}
\usepackage[dvipsnames]{xcolor}
\newcommand{\fnd}[2]{\frac{\textstyle #1}{\textstyle #2}}
\newcommand{\fndrs}[4]{\fnd{\raisebox{#1}{$#2$}}{\raisebox{#3}{$#4$}}}
\newcommand{\Real}[1]{\Re {\it e}\left(#1 \right)}
\newcommand{\Imag}[1]{\Im {\it m}(#1 )}
\newcommand{\dissum}[2]{\displaystyle \sum_{#1}^{#2}}

\begin{document}
\title{Modeling two-boson mass distributions,\\ E(38 MeV) and Z(57.5 GeV)%
\thanks{Presented at Excited QCD 2016,
Costa da Caparica (Portugal), 6-12 March 2016}%
}
\author{Eef van Beveren
\address{Centro de F\'{\i}sica da UC,
Departamento de F\'{\i}sica,\\ Universidade de Coimbra,
P-3004-516 Coimbra, Portugal}
\\ [10pt]
George Rupp
\address{Centro de F\'{\i}sica e Engenharia de Materiais Avan\c{c}ados,\\
Instituto Superior T\'{e}cnico,
Universidade de Lisboa,
P-1049-001 Lisboa, Portugal}
\\ [10pt]
Susana Coito
\address{Institute of Modern Physics, CAS, Lanzhou 730000, China}
}
\maketitle
\begin{abstract}
Besides general features of the Resonance Spectrum Expansion
for two-boson mass distributions,
experimental results are discussed.\\
Furthermore, E(38 MeV) and Z(57.5 GeV) are highlighted.
\end{abstract}

\section{Introduction}

The Resonance Spectrum Expansion (RSE) for two-boson mass distributions
is a general expression for the two-boson scattering amplitude
in the presence of an infinite tower of s-channel resonances.
A complete derivation of the RSE formula at elementary level
can be found in Ref.~\cite{IJTPGTNO11p179}.
Here, we will take a shortcut via the Breit-Wigner expression
for the two-boson scattering amplitude $T(\sqrt{s})$
in the presence of one resonance at $\sqrt{s}=M$, given by
\begin{equation}
T\left(\sqrt{s}\right)=\fndrs{2pt}{\lambda^{2}\Imag{F(s)}}{-2pt}
{\sqrt{s}-M+\lambda^{2}F(s)}
=\fndrs{2pt}{\fnd{\lambda^{2}\Imag{F(s)}}{\sqrt{s}-M}}{-2pt}
{1+\fnd{\lambda^{2}F(s)}{\sqrt{s}-M}}
\;\;\; ,
\label{BW}
\end{equation}
where $F(s)$ represents the two-boson loop function
and $\lambda$ the coupling strength of the three-boson vertex.
Notice that the peak $M_{R}$ of the resonance enhancement
comes at $M_{R}=M-\lambda^{2}\Real{F(s)}$,
whereas its width is given by $\Gamma_{R}\approx -2\lambda^{2}\Imag{F(s)}$.

In the presence of an infinite tower of s-channel resonances,
$M_{0}$, $M_{1}$, $M_{2}$, $\dots$,
the two-boson scattering amplitude takes the form
\begin{equation}
T\left(\sqrt{s}\right)=
\fndrs{2pt}{\lambda^{2}\Imag{F(s)}
\dissum{n=0}{\infty}\fnd{g_{n}^{2}}{\sqrt{s}-M_{n}}}{-2pt}
{1+\lambda^{2}F(s)\dissum{n=0}{\infty}\fnd{g_{n}^{2}}{\sqrt{s}-M_{n}}}
\;\;\; ,
\label{RSE}
\end{equation}
where $g_{n}$ represents the relative coupling of the two bosons
to the $n$-th resonance \cite{ZPC21p291}.
In a multi-channel description the ingredients of formula (\ref{RSE})
turn into matrices.
Properties of the scattering amplitude (\ref{RSE}) have been studied
in a series of papers
(See {\it e.g.} \cite{APPS8p139} and references therein).

When the overall coupling $\lambda$ is small,
the corresponding mass distributions show narrow resonance peaks
near the $M_{n=0,\, 1,\, 2,\,\dots}$ masses (seeds).
However, when $\lambda$ takes realistic values
the resonance peaks become broader and shift away from the seed masses
yielding the experimentally observed resonance
central masses and widths.
Seeds represent the underlying quark-antiquark spectrum
which is hence not identical
to the observed resonance central mass spectrum.

It is opportune to mention here that formula (\ref{RSE})
also applies to resonances below the strong thresholds.
For example, it predicts as well the charmonium bound states
$J/\psi$(1S) and $\psi$(2S) below the $D\bar{D}$ threshold
as the resonances above that threshold \cite{PRD21p772,CNPC35p319}.
This property is due to full analyticity of formula (\ref{RSE})
in the total invariant mass.

Observable quantum numbers, like total angular momentum, parity
and C-parity are respected for $\lambda\neq 0$,
but, internal quantum numbers, like radial excitation and
relative angular momentum, are not.
Hence, the resulting resonance "states"
do not have pure radial excitation or relative angular momentum.
For example, $J^{PC}=1^{--}$ charm-anticharm vector bosons,
which have seeds with internal angular momenta
$\ell =0$ (S) and $\ell =2$ (D),
turn into charmonium resonances with mixed configurations
of S and D-states.
Moreover, the dominantly D-states almost decouple from scattering,
leading to narrow resonances which hardly shift away
from the seed masses, whereas the dominantly S-states couple
more strongly to scattering, leading to broader resonances
which shift hundreds of MeV's away from the seed masses.
As a consequence, one can almost identify the seed spectrum
with the dominantly D-states.
Unfortunately experimental observations are lacking.

A further consequence of realistic values for $\lambda$
is the appearance of dynamically generated resonances
which do not have a direct relation to the seeds.
Examples are the low-lying scalar resonances \cite{ZPC30p615}
and the $D_{s0}^{\ast}$(2317) resonance \cite{PRL91p012003}.

In Ref.~\cite{AP323p1215} an expression has been deduced
for production of boson pairs which relates the production ($P$)
and scattering amplitudes according to
\begin{equation}
P\left(\sqrt{s}\right)=
\Imag{Z(s)}+T\left(\sqrt{s}\right)Z(s)
\;\;\; ,
\label{PRSE}
\end{equation}
where $Z(s)$ is a purely kinematic expression
which contains no singularities.
Resonance poles of the scattering amplitude (\ref{RSE}) determine fully
the singularity structure of the production amplitude (\ref{PRSE}).
Consequently, resonances in scattering also show up in production.
But, the shape of $\Imag{Z(s)}$ is such that,
in the ideal case of no further nearby thresholds,
it rises sharply just above threshold.
For larger invariant masses $\Imag{Z(s)}$ first reaches a maximum
and then falls off exponentially.
As a consequence, production amplitudes show non-resonant
yet resonant-like enhancements just above threshold \cite{PRD80p074001}.
From the invariant mass at its peak one can estimate \cite{APPS8p145}
the interaction distance $a$ by
\begin{equation}
2(pa)^{2}\approx 1
\;\;\; ,\;\;\;
\sqrt{s}=\sqrt{p^{2}+m_{1}^{2}}+\sqrt{p^{2}+m_{2}^{2}}
\;\;\; ,
\label{idist}
\end{equation}
where $m_{1,2}$ represent the masses of the produced bosons.

\section{Exotics}

With the formalism developed in Ref.~\cite{ZPC21p291}
one not only can determine
the relative coupling constants $g_{n=0,\, 1,\, 2,\,\dots}$
of the seeds to the various two-boson channels,
but also the number of two-boson channels
which couple to each of the seeds.
This number grows rapidly with radial excitation.
As a consequence higher radial excitations
couple much more weakly to a given two-boson channel
than the ground state.
The relative coupling squared of the $n$-th radial excitation
to a given two-boson channel drops proportionally
to the $n$-th power of 4, times a polynomial in $n$,
while the number of two-boson channels
which couple to the $n$-th radial excitation grows correspondingly.
Most of those channels are closed for decay
since the masses of the two bosons are too high.
Nevertheless, the fifth or the sixth radial excitation
of a certain flavor-antiflavor configuration
couples more weakly to a given two-boson channel than to weak decay.

As an example, suppose that in experiment one measures
a weak decay channel $J/\psi\pi^{+}$ \cite{PRL110p252001}
near the sixth or seventh radial excitation of the $c\bar{s}$ system.
Then one obtains a resonance signal for the excitation.
Its mass is given by the mass of the seed and a hadronic shift,
whereas its width is determined
by all the open and closed strong two-boson channels.
As long as the open candidates are not yet looked for in experiment,
one may prematurely conclude that the signal stems
from an exotic quark system \cite{Hadron2013p037}.
However, only a full inspection
of all of the many possible strong decay channels
for the corresponding $c\bar{s}$ system can resolve this and,
since the various
$c\bar{u}+u\bar{s}$, $c\bar{d}+d\bar{s}$ and $c\bar{s}+s\bar{s}$
two-boson channels and their excitations
couple very weakly to the sixth or seventh radial excitation
of the $c\bar{s}$ system, that may need some statistics.
Moreover, the reconstruction of those channels
out of kaons, pions and electron-positron pairs
constitutes quite a larger challenge for experiment
than measuring the rather easy weak $J/\psi\pi^{+}$ channel.
We are thus still far away from the discovery
of exotic quark configurations.

\section{E(38 MeV) scalar boson}

In Refs.~\cite{ARXIV12021739,EPJWC95p02007}
a variety of indications were presented
of the possible existence of a light boson with a mass of about 38 MeV.
These indications amounted to a series of low-statistics observations
all pointing in the same direction, and one high-statistics observation,
which might be interpreted as the discovery of the E(38 MeV).

About three decades ago it could be observed
from the results of Ref.~\cite{PRD27p1527}
that $^{3}P_{0}$ pair creation is associated with a light quantum.
Nevertheless, values of 30--40 MeV for its flavor-independent mass
did not seem to bear any relation
to an observed quantity for strong interactions.
However, in Refs.~\cite{ARXIV12021739,ARXIV11021863}
we have presented experimental evidence
for the possible existence of a quantum with a mass of about 38 MeV,
which in light of its relation to the $^{3}P_{0}$ mechanism
we suppose to mediate quark-pair creation.
Moreover, its scalar properties make it a perfect candidate
for the quantum associated with the scalar field for confinement
\cite{NCA80p401}.

\section{Weak substructure and the Z(57.5 GeV)}

In Refs.~\cite{ARXIV13047711,ARXIV14114151}
we have indicated the possible existence of substructure
in the weak sector, based on the observation
that recurrences may exist for the $Z$ boson.
The corresponding data do not have sufficient statistics
to yet conclude the existence of Weak substructure,
except perhaps for a clear dip at about 115 GeV
in diphoton, four-lepton, $\mu\mu$ and $\tau\tau$
invariant-mass distributions.
The latter structure indicates the possible opening of
a two-particle threshold, probably pseudo-scalar partners
of the $Z$ boson with masses of about 57.5 GeV.
Further possible recurrences of the $Z$ boson,
observed by us at 210 and 240 GeV,
certainly need a lot more statistics.

Composite heavy gauge bosons and their spin-zero partners,
the latter with a mass in the range 50--60 GeV,
were considered long ago \cite{PLB135p313}
and studied in numerous works.
To date, no experimental evidence of their existence has been reported.
However, if a pseudo-scalar partner
of the $Z$ boson with mass of about 57.5 GeV exists
and, consequently, part of the structure
observed in the mass interval 115--135 GeV
is interpreted as a threshold enhancement,
then it must be possible to verify its existence at LHC,
for example in four-photon events.

More recently the interest in weak substructure has revived
\cite{ARXIV12074387,ARXIV12105462,ARXIV13076400,ARXIV13040255,PRD90p035012}.
Most popular among the proposed models is the Technicolor Model (TC)
\cite{PRD20p2619} for which one expects QCD-like dynamics
but much stronger.
From the structure of the threshold enhancement above 115 GeV,
we deduced an interaction distance of the order of 0.008 fm
\cite{ARXIV14114151}.
Now, from QCD we have learned that self--interactions
lead to an appreciable contribution to the masses of resonances.
Hence, for yet much stronger dynamics we must expect
that the masses of resonances are basically determined
by the self--interactions and not so much
by the masses and binding forces of the constituents.
This has, indeed, been recognized in Ref.~\cite{PRD90p035012}
where, in a perturbative fashion, the mass of the TC scalar resonance
is lowered by several hundreds of GeVs.
However, as we have argued that already for QCD unquenching
should be incorporated beyond perturbative contributions,
we assume that for weak substructure it is indispensable to do so.
This, furthermore, implies that the corresponding spectrum
will also contain dynamically generated resonances
and may even be dominated by such poles,
rather than by those which stem from confinement.

\section{Conclusions}

Modeling the dynamics of strong interactions is useful.
However, it must be accompanied by
the study of scattering and production \cite{PLB747p410}
in the presence of towers of resonances,
not just isolated enhancements.
Experiment, unfortunately, does not yet provide the necessary statistics
to confront model results with measured cross sections.

\newcommand{\pubprt}[4]{#1 {\bf #2}, #3 (#4)}
\newcommand{\ertbid}[4]{[Erratum-ibid.~#1 {\bf #2}, #3 (#4)]}
\def\AP{{\it Ann.\ Phys.}}
\def\APPS{{\it Acta Phys.\ Polon.\ Supp.}}
\def\CNPC{{\it Chin.\ Phys.\ C}}
\def\EPJWC{{\it Eur.\ Phys.\ J.\ Web of Conf.}}
\def\IJTPGTNO{{\it Int.\ J.\ Theor.\ Phys.\ Group Theor.\ Nonlin.\ Opt.}}
\def\NCA{{\it Nuovo Cim.\ A}}
\def\PLB{{\it Phys.\ Lett.\ B}}
\def\PRD{{\it Phys.\ Rev.\ D}}
\def\PRL{{\it Phys.\ Rev.\ Lett.}}
\def\ZPC{{\it Z.\ Phys.\ C}}


\begin{thebibliography}{27}
\bibitem{IJTPGTNO11p179}
E.~van Beveren and G.~Rupp,
\pubprt{\IJTPGTNO}{11}{179}{2006}
[arXiv:hep-ph/0304105].

\bibitem{ZPC21p291}
E.~van Beveren,
\pubprt{\ZPC}{21}{291}{1984}
[arXiv:hep-ph/0602246].

\bibitem{APPS8p139}
G.~Rupp, E.~van Beveren and S.~Coito,
\pubprt{\APPS}{8}{139}{2015}
[arXiv:1502.05250].

\bibitem{PRD21p772}
E.~van Beveren, C.~Dullemond, and G.~Rupp,
\pubprt{\PRD}{21}{772}{1980}
\ertbid{\ D}{22}{787}{1980}.

\bibitem{CNPC35p319}
E.~van Beveren and G.~Rupp,
\pubprt{\CNPC}{35}{319}{2011}
[arXiv:1004.4368].

\bibitem{ZPC30p615}
E.~van Beveren, T.~A.~Rij\-ken, K.~Metzger, C.~Dullemond, G.~Rupp and
J.~E.~Ribeiro,
\pubprt{\ZPC}{30}{615}{1986}
[arXiv:0710.4067].

\bibitem{PRL91p012003}
E.~van Beveren and G.~Rupp,
\pubprt{\PRL}{91}{012003}{2003}
[arXiv:hep-ph/0305035].

\bibitem{AP323p1215}
E.~van Beveren and G.~Rupp,
\pubprt{\AP}{323}{1215}{2008}
[arXiv:0706.4119].

\bibitem{PRD80p074001}
E.~van Beveren and G.~Rupp,
\pubprt{\PRD}{80}{074001}{2009}
[arXiv:0908.0242].

\bibitem{APPS8p145}
E.~van Beveren, G.~Rupp and S.~Coito,
\pubprt{\APPS}{8}{145}{2015}
[arXiv:1502.04862].

\bibitem{PRL110p252001}
M.~Ablikim {\it et al.}  [BESIII Collaboration],
\pubprt{\PRL}{110}{252001}{2013}
[arXiv:1303.5949].

\bibitem{Hadron2013p037}
M.~Nielsen,
PoS Hadron {\bf 2013}, 037 (2013).

\bibitem{ARXIV12021739}
E.~van Beveren and G.~Rupp,
arXiv:1202.1739.

\bibitem{EPJWC95p02007}
E.~van Beveren, S.~Coito and G.~Rupp,
\pubprt{\EPJWC}{95}{02007}{2015}
[arXiv:1411.4151].

\bibitem{PRD27p1527}
E.~van Beveren, G.~Rupp, T.~A.~Rij\-ken, and C.~Dullemond,
\pubprt{\PRD}{27}{1527}{1983}.

\bibitem{ARXIV11021863}
E.~van Beveren and G.~Rupp,
arXiv:1102.1863.

\bibitem{NCA80p401}
C.~Dullemond, T.~A.~Rij\-ken and E.~van Beveren,
\pubprt{\NCA}{80}{401}{1984}.

\bibitem{ARXIV13047711}
E.~van Beveren, S.~Coito and G.~Rupp,
arXiv:1304.7711.

\bibitem{ARXIV14114151}
E.~van Beveren, S.~Coito and G.~Rupp,
arXiv:1411.4151.

\bibitem{PLB135p313}
U.~Baur, H.~Fritzsch and H.~Faissner,
\pubprt{\PLB}{135}{313}{1984}.

\bibitem{ARXIV12074387}
T.~Matsushima,
arXiv:1207.4387.

\bibitem{ARXIV12105462}
E.~Eichten, K.~Lane and A.~Martin,
arXiv:1210.5462.

\bibitem{ARXIV13076400}
H.~Fritzsch,
arXiv:1307.6400.

\bibitem{ARXIV13040255}
E.~H.~Simmons, A.~Atre, R.~S.~Chivukula, P.~Ittisamai,
N.~Vignaroli, A.~Farzinnia and R.~Foadi,
contribution to SCGT12 "KMI-GCOE workshop on strong coupling gauge theories
in the LHC perspective", 4-7 Dec. 2012, Nagoya University,
arXiv:1304.0255 [hep-ph].

\bibitem{PRD90p035012}
A.~Belyaev, M.~S.~Brown, R.~Foadi and M.~T.~Frandsen
\pubprt{\PRD}{90}{035012}{2014}
[arXiv:1309.2097].

\bibitem{PRD20p2619}
L.~Susskind,
\pubprt{\PRD}{20}{2619}{1979}.

\bibitem{PLB747p410}
A.~P.~Szczepaniak,
\pubprt{\PLB}{747}{410}{2015}
[arXiv:1501.01691].

\end{thebibliography}
\end{document}